# Transformative Technology for FLASH Radiation Therapy: A Snowmass 2021 White Paper

S. Boucher[1], E. Esarey[2], C.G.R. Geddes[2], C. Johnstone[3], S. Kutsaev[1], B.W. Loo Jr.[4], F. Méot[5], B. Mustapha[6], K. Nakamura[2], E. Nanni[9], L. Obst-Huebl[2], S.E. Sampayan[7], C. Schroeder[2], R. Schulte[8], K. Sheng[9], A. Snijders[2], E. Snively[10], S.G. Tantawi[10], J. van Tilborg[2]

[1] RadiaBeam Technologies, LLC, Santa Monica, CA 90404, USA

[2] Lawrence Berkeley National Laboratory, CA, 94720, USA

[3] Fermi National Accelerator Laboratory, Batavia, IL, 60510, USA

[4] Stanford University School of Medicine, Stanford, CA 94305, USA

[5] Brookhaven National Laboratory, Upton, NY 11973, USA

[6] Argonne National Laboratory, Lemont, IL 60439, USA

[7] Lawrence Livermore National Laboratory, Livermore, CA 94551, USA; and Opcondys, Inc., Manteca, CA 95336, USA

[8] Loma Linda University, Loma Linda, CA 92350, USA

[9] University of California, Los Angeles, CA 90095, USA

[10] SLAC National Accelerator Laboratory, Menlo Park, CA 94025, USA

**Keywords**: Accelerators, FLASH effect, Radiation Therapy

**Abstract**

Conventional cancer therapies include surgery, radiation therapy, chemotherapy, and, more recently, immunotherapy. These modalities are often combined to improve the therapeutic index. The general concept of radiation therapy is to increase the therapeutic index by creating a physical dose differential between tumors and normal tissues through precision dose targeting, image guidance, and high radiation beams that deliver radiation dose with high conformality, e.g., protons and ions. However, treatment and cure are still limited by normal tissue radiation toxicity, with many patients experiencing acute and long-term side effects. Recently, however, a fundamentally different paradigm for increasing the therapeutic index of radiation therapy has emerged, supported by preclinical research, and based on the FLASH radiation effect. FLASH radiation therapy (FLASH-RT) is an ultra-high dose-rate delivery of a therapeutic radiation dose within a fraction of a second. Experimental studies have shown that normal tissues seem to be universally spared at these high dose rates, whereas tumors are not. The dose delivery conditions are not yet fully characterized. Still, it is currently estimated that large doses ($\geq$10 Gy) delivered in 200 ms or less produce normal tissue sparing effects yet effectively kill tumor cells. There is a great opportunity, but also many technical challenges, for the accelerator community to create the required dose rates with novel and compact accelerators to ensure the safe delivery of FLASH radiation beams.

.

# 1 Introduction

Radiation therapy is a dynamic research field driven by new technology developments. An exciting recent discovery is the sparing of normal (non-tumor) tissues when irradiated with ultra-high dose rates, but tumors are not spared when irradiated with the same radiation field, as first reported by Favaudon and colleagues in 2014 [1]. This phenomenon, which is now called the FLASH effect, opens up a potential new modality in radiation therapy (RT). Tissue sparing means that much higher radiation doses than conventional ones are tolerated increasing the potential for a cure with an accompanying reduction in side effects.

Many preclinical and first clinical results indicate a dramatic reduction of toxicity response at FLASH-RT dose rates compared to conventional dose rates. The first human patient treated with FLASH 2019 (a patient with recurrent cutaneous T-cell lymphoma) had a complete response with minimal skin toxicity despite being previously and repeatedly treated with conventional photon and electron RT at the tumor site [2]. FLASH-RT produced a complete response and was exceptionally tolerated even after multiple non-FLASH skin irradiations had produced significant radiation damage from both photons and electrons.

Most FLASH experiments and preclinical studies have been performed with electrons and only very few with protons. However, conventional radiotherapeutic electron beams (4-25 MeV) cannot penetrate enough to treat deep-seated tumors and are thus not likely to be widely used clinically for FLASH-RT. The distal dose fall-off of electrons in tissue is quite shallow as electrons are scattered and have considerable energy straggling, especially at low energies. Protons are currently the most commonly used heavy-charged particle in RT with a Bragg peak dose and finite range advantage. They are also beginning to be used for the first treatment planning studies for clinical trials with FLASH-RT [3, 4]. Heavy ions, like carbon and helium, are not widely used despite a potentially much greater therapeutic effect due to high linear energy transfer (LET). The high associated costs of ion accelerator and gantry technologies have prevented widespread use. However, several carbon ion facilities are operational in Europe and Asia, particularly Japan.

Considerable research and development in this area is essential to optimize and clinically realize the curative potential of FLASH-RT with different radiation modalities. Currently, electron FLASH studies are performed using 4-6 MeV electron beams from modified clinical linacs and provide the strongest, most consistent preclinical evidence for the FLASH effect. Experimental high dose rate photon beams have been formed using synchrotron radiation and keV X-rays from a tube (very early FLASH-RT studies) with mixed results for the FLASH effect. The FLASH effect has also been observed with protons using shoot-through beams from clinical CW or iso-cyclotrons. In shoot-through beams, the beam is not energy degraded, so the proton energy ranges from 230-250 MeV, i.e., the highest available proton energy with these cyclotrons. Once energy degraders are introduced into the beam to create lower energy proton beams, FLASH intensities cannot be achieved. Synchrotrons, even the rapid cycling 15 Hz ion synchrotron being developed at BNL, cannot produce the intense ion beams required for a clinical application of FLASH-RT– and only a very small volume can be irradiated at the cycle time of the synchrotron.

## 2  Beam conditions for the FLASH Effect

Preclinical electron studies have generated a self-consistent set of general beam conditions for FLASH as >40Gy/sec, >10 Gy delivered in <250 ms, with an instantaneous dose rate >$10^6$ Gy/sec instantaneous dose rate during a microsecond beam pulse [5-7]. The instantaneous dose rate requirement appears to be of particular significance as it has been observed across all radiation modalities, as summarized in Figure 1. The two proton outliers may reflect a difference between the beam structure of the quasi-continuous wave (CW) proton therapy beam and the 100-300 Hz pulsed electron beam, which has an RF bunch microstructure. The RF GHz bunch microstructure of the electron beam is approximately two orders of magnitude shorter than the MHz RF bunch structure of the proton beam. How these different time structures influence the observation of a FLASH effect is unclear as systematic studies of pulse duration and frequency in different biological systems have yet to be done.

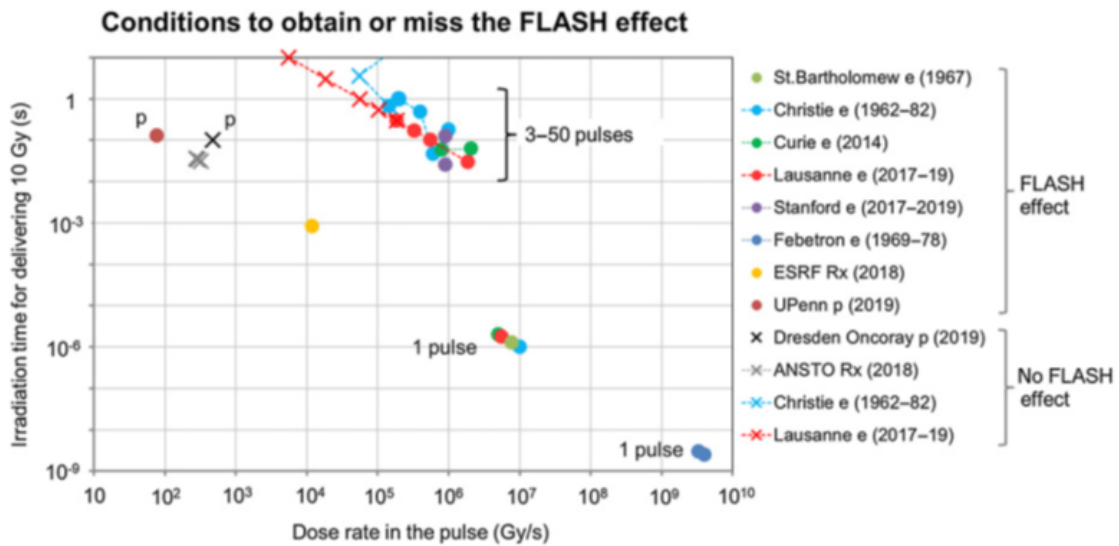

**Figure 1.** Summary of preclinical studies at different accelerator facilities with different radiation types (right panel). Note the irradiation time for delivering 10 Gy on the vertical axis and the instantaneous dose rate on the horizontal axis. The FLASH effect begins at ~$10^5$ Gy/sec dose rate during the pulse and a < 1 sec delivery time, where 0.1 s duration and 106/s instantaneous dose rate are firmly in the FLASH region. FLASH effects were seen with nanosecond pulses and more than $10^9$ Gy/sec. Reproduced from Montay-Gruel P et al. [8]

## 2.1 FLASH with electron beams

Highlights of observed FLASH effects from <10 MeV electron beams using 100 Hz clinical linacs are listed below with references. Beam properties from a compilation of studies observing FLASH electron beam conditions are summarized in Table 1.

- Study of pulmonary fibrosis from irradiation of the lung [1]
  - Severe to moderate for conventional average dose rate of 0.03 Gy/sec, 17 Gy total dose
  - For an average dose rate of 40-60 Gy/sec, equivalent fibrosis occurred at 30 Gy total dose
- Study of neurocognitive impairment in mice from brain irradiation [5]
  - Severe neurocognitive degeneration at an average dose rate of 0.1 Gy/sec, 10 Gy total dose
  - Reduced impairment starts at 30 Gy/sec with no neurocognitive decline at 100 Gy/sec average dose rate for 10 Gy!
- Skin irradiation (mini-pig) [6]
  - Fibrosis and necrotic lesions were observed at an average dose rate of 0.08 Gy/sec (22-37 Gy total dose)
  - Only mild depigmentation at an average dose rate of 300 Gy/sec (22-37 total dose)!

**Table 1.** Preclinical FLASH properties relevant to a clinical application of FLASH

| Electron Beam | Min for observed FLASH | Optimal for FLASH |
|---|---|---|
| Average Dose Rate | 40 Gy/sec | 100 Gy/sec |
| Instantaneous Dose Rate | $\sim 10^5$ Gy/sec | $\geq 10^6$ Gy/sec |

| Total Dose | 10 Gy | ≥10 Gy – tissue dependent |
|---|---|---|
| Delivery Time for 10 Gy | nanoseconds | <1 second |

Additionally, there are proposals for Very High Energy Electron beams (VHEE) for a more penetrating clinical electron beam. Tumor depths of 30 cm require 200-250 MeV electrons. Treatment models have predicted that to deliver 10 Gy/sec requires $10^{11}$ e/sec for a 200 MeV, σ=1.5 mm Gaussian beam (D. Bartosik, private communication). Although not discussed in depth here, there are existing electron accelerator facilities that could be used for FLASH studies. The 5Hz FAST SRF Linac at Fermilab produces 50 and 300 MeV beams capable of delivering up to 1000 Gy/pulse and a $10^6$ Gy/sec instantaneous rate based on the above simulation. Clinical scanning capability requires a faster duty cycle of ~100 Hz or higher. The CBETA CW recirculating energy recovery linac at Cornell University produces a 150 MeV beam and can scan $10^6$ Gy/sec at 200 cm/ms (beyond current transverse scanning capability). The size of the ring can be dramatically reduced by replacing the weak permanent magnets in the arcs, and a miniaturized version is under conceptual design.

### 2.2   FLASH with photon beams

Photon dose rates produced using clinical electron linacs are too low, and no FLASH effects have been observed. FLASH dose rates can be produced using a high-intensity SRF electron linac with a tungsten target, and a significant FLASH effect was observed for lungs and other tissues. An early 1969 study using high-dose-rate X-rays delivered with nanosecond pulses to mammalian cells in vitro showed sparing effects compared to conventional dose rate [9]. A recent Monte Carlo

Study [10] suggests that FLASH with X-ray tubes may be possible, and a first experimental demonstration of the FLASH effect with X-rays was reported by Gao et al. [11]

High dose rates of photons can also be produced using light-source synchrotrons, notably Synchrotron Broad-Beam Radiation therapy (SBBR) and Microbeam Radiation therapy (MRT) which generates a grid of "pencil" photon beams1. For the SBBR, one study did not show a FLASH effect (37 – 41 Gy/sec, 4-28 Gy) [11]. Still, a second study involving mouse-brain irradiation (37 Gy/sec, 10 Gy) showed significant cognitive sparing similar to the experience with electron FLASH [12]. The main difference between the two studies was that the second had a vertical beam size that was 20 times smaller than the first. The MRT preclinical studies generate a parallel beam array, with 25-100 µm wide peaks and 100-400 µm wide valleys. The peak average dose rate was ~300 Gy/sec, but the valley average dose rate was a lower factor of ~30, and the valleys were a strong indicator of toxicity. The low valley dose rates were conjectured to be the reason for the absence of the FLASH effect.

### 2.3  FLASH with proton and ion beams

Although pulsed FLASH has been proposed using large synchrotrons and fast extraction, the dose volume remains small and scanning problematic as the beam would likely be re-positioned between spills which makes the treatment time likely incompatible with FLASH conditions. FLASH requires ultra-high repetition rates like the electron linacs or CW beam as output from 230-250 MeV proton therapy iso-cyclotrons. The FLASH effect is only observed with shoot-through (non-degraded) beams that can achieve FLASH intensities. This high-energy beam places the Bragg peak behind the targeted area. To date, there are relatively few proton FLASH studies with mixed results [14,15]. However, many more preclinical studies are underway. Individual RF (MHz) proton bunch structure may be important for proton FLASH (see Discussion below); proton RF "bunches' are fractions of a microsecond, and electron RF bunches are fractions of a nanosecond. Further, CW proton beams are "quasi-continuous" whereas 100-300 Hz electron linacs produce an ~microsecond "macro-pulse" containing many RF bunches and the "instantaneous" dose rate is averaged over the macro-pulse. For proton FLASH, it can be hypothesized that the instantaneous dose rate of $10^5 - 10^6$ Gy/sec must be achieved across the RF bunch pulse. For pulsed electron beams, the instantaneous dose rate is integrated over the microsecond individual macro-pulse, which repeats at the 100-300 Hz rate of the pulsed clinical linac., as shown in Figure 2. The usually quoted 100 ms treatment time as optimal for FLASH may not apply to quasi-continuous proton beams or CW SRF electron linacs.

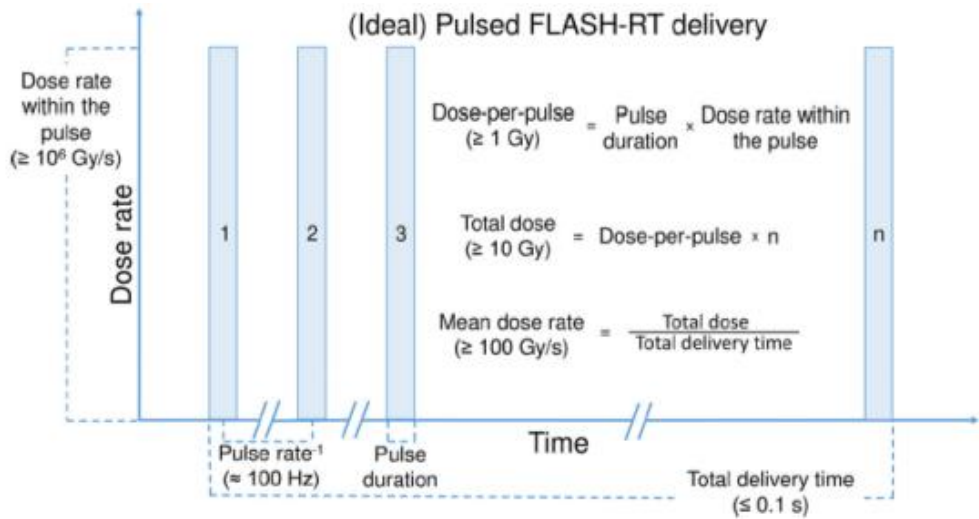

**Figure 2.** Schematic view of pulsed beam delivery inducing the FLASH effect. Reproduced from Montay-Wilson JD et al. [7]

FLASH intensities have been demonstrated using proton-therapy iso-cyclotrons (cyclotrons that produce continuous rather than pulsed beams) and are currently the only proton facilities capable of participating in FLASH preclinical patient trials. In regard to ion therapy, currently only synchrotrons are available and synchrotrons, even the rapid cycling 15 Hz ion synchrotron, are not capable of producing the intense ion beams required for FLASH radiotherapy for a clinical application. In addition to providing intense ion beams, FLASH radiotherapy studies need to be extended to incorporate the Bragg peak, range dependencies, and dosimetry into a broader research initiative. Since there are innovative accelerator technologies under development that may provide FLASH intensities for protons and ions, beam conditions can be proposed for the FLASH effect based on RBEs and LETs between the different particle beams. These extrapolations are presented in Table 2 and compared with a more conventional dose rate and fraction.

**Table 2.** Dose delivery requirements for FLASH for protons and ions extrapolated from electron FLASH studies

| Dose Delivery Mode | Protons | Helium | Carbon |
|---|---|---|---|
| <u>Conventional Fraction</u>: 2.6 Gy/fraction<br><br>Delivery time: 100 sec | $2.6 \times 10^9$ p/sec<br><br>0.4 nA | $6.5 \times 10^8$ He/sec<br><br>0.2 nA | $2.2 \times 10^8$ C/sec<br><br>0.2 nA |
| <u>FLASH</u>: $\geq 10$ Gy/fraction<br><br>Delivery Time: 100 msec | $1 \times 10^{13}$ p/sec<br><br>1.6 $\mu$A | $2.5 \times 10^{12}$ He/sec<br><br>0.8 $\mu$A | $0.8 \times 10^{12}$ C/sec<br><br>0.8 $\mu$A |

### 2.4  Discussion

Most preclinical studies published to date have not been systematically coordinated around beam parameters, including mean and instantaneous dose rate, total dose, pulse structure, fractionation, and radiation type. Although there is a wide range of dose rates, some observations can be made nonetheless, although these may apply only to electron FLASH.[5,7,12].

- FLASH effects – general
  - Appear at average dose rates >30 Gy/sec, apparent optimal at 100 Gy/sec
  - FLASH effect likely highly tissue dependent
- Dependence on the beam structure and uniformity in dose deposition
  - Beam delivery
  - Typical dose delivery time for a consistent (electron) FLASH effect is~100 milliseconds (best <250 ms)
  - MOST positive FLASH studies used a pulsed clinical electron linac with a beam pulse length of ~microseconds and a repetition rate of 100-400 Hz.
  - Instantaneous (within the pulse) minimum FLASH dose rate is $10^6$ Gy/sec (again, a characteristic of clinical electron linacs).
- Dosimetry and treatment planning questions
  - Observed volumetric dose deposition dependence
  - Low dose-rate areas not tolerated during FLASH – toxicity reappears2
  - Bragg peak and pencil beam scanning questions: do distal edge and penumbra effects and associated lower-dose rate beam "halos" create a problem?
  - Can a relatively large target volume be uniformly irradiated by fractionated FLASH compatible deliveries over a longer time frame?
  - Instantaneous FLASH dose rate and delivery time for 10 Gy -is it consistent for all radiation types

## 3  Emerging accelerator technologies for FLASH-RT

### 3.1  High-gradient ion linacs for FLASH-RT with ion beams

Synchrotrons are used for ion beam therapy, while cyclotrons are mainly used for proton therapy. Until recently, linacs were not seriously considered for ion beam therapy due to the required accelerator length and extended footprint. With the recent developments of high-frequency high-gradient accelerating copper structures, more compact linacs are being proposed for protons and ions. These structures should be capable of delivering FLASH beam intensities.

### 3.1.1 Linacs for ion beam therapy

Being a single-pass machine, a pulsed linear accelerator (linac) is capable of adjusting the pulse repetition rate and the beam energy hundreds of times per second (~ 200 Hz). This much-desired flexibility in beam tuning enables fast and efficient beam scanning to allow 3D dose painting, as well as real-time image-guided range calculation and targeting of moving targets. By changing the pulse repetition rate, the beam intensity could be adjusted up to $10^9$ ions per second ($10^{10}$ for protons), typically needed for therapy. For carbon ions, the energy could be changed continuously up to the full energy of 430 MeV/u required to penetrate the depth of a human body, which is equivalent to 30 cm of water. In addition, the beam quality from a single-pass full energy linac is generally better than other systems that may require energy degraders or multi-turn acceleration.

Linacs have already been proposed for proton [16] and carbon beam therapy [17], but no all-linac-based facilities exist. This is due to the length and space required for the linac, which has limited its deployment in a hospital or other clinical setting. Using traditional accelerator technology, a linac would be hundreds of meters long, and this is the main reason why synchrotrons are currently dominating the field of ion beam therapy. The beam delivery from a linac will be similar to synchrotron beam delivery through fixed beamlines or gantry systems. However, the superior beam quality of the linac enables much smaller magnets and, therefore, more compact gantries.

### 3.1.2   The ACCIL ion linac: General and FLASH capabilities

The Advanced Compact Carbon Ion Linac (ACCIL) is the most compact full-energy carbon ion linac proposed for therapy [18]. In Europe, there are proposals for a combined cyclotron and linac (cyclinac) and an all-linac for carbon beams [19] [not really clear to me how Ref.3.1.2 fits in here], in addition to the ongoing LIGHT project for a proton therapy linac [19] [same here]. ACCIL is designed to deliver a full energy of 450 MeV/u, which exceeds the maximum energy required for carbon ion therapy. It is also capable of accelerating protons and many other ion beams to the same energy per nucleon. Figure 3 presents a schematic layout of the ACCIL design. The system is about 45 m long but could, in principle, be folded into two 25 m long sections.

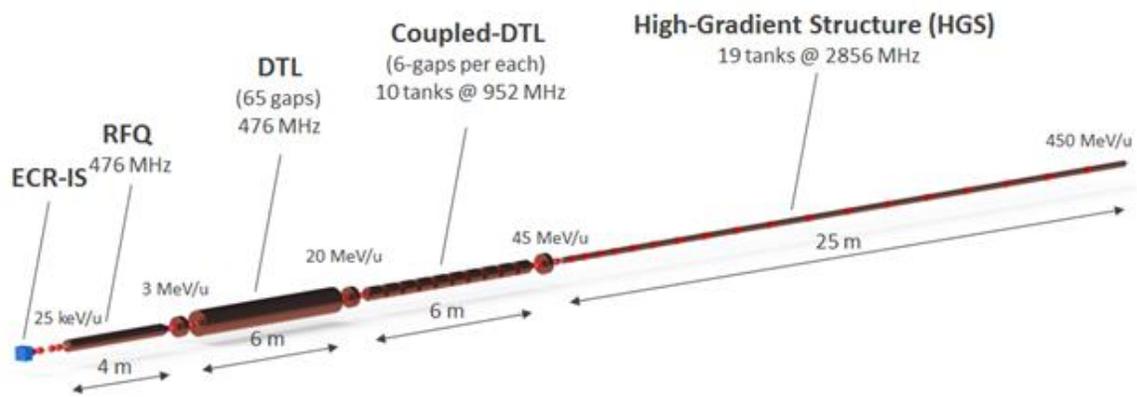

**Figure 3.** Layout of the proposed ACCIL design.

The linac is comprised of an electron cyclotron resonance ion source, followed by a radio-frequency quadrupole accelerating the carbon beam to 3 MeV/u, which is further accelerated in a drift tube linac (DTL), then in a coupled DTL linac up to 45 MeV/u. The essential features to achieve compactness in the ACCIL design are high-gradient structures, each capable of delivering 50 MV/m, that accelerate the beam to the full energy of 450 MeV/u in ~ 25 meters.

The main advantages of ACCIL are fast pulse-to-pulse beam energy change and ion beam switching capabilities. Different ion sources could be used in the front-end to allow fast beam switching between different ion species. The delivered beam intensity could also be controlled by adjusting the pulse length at the source or changing the pulse repetition rate, typically from 100 to 400 pulses per second (pps) are possible, and R&D for accelerating structures capable of operating at 1000 pps is ongoing. Ultimately, the tuning flexibility of the ACCIL design will allow fast and effective variable-energy and intensity-modulated multi-ion beam therapy.

ACCIL is capable of accelerating a variety of ion beams from proton to neon, up to a maximum energy of 450 MeV/u. At this energy, ions lighter than carbon, including protons and helium ions, have ranges exceeding the depth of the human body and could therefore be used for imaging such as proton tomography. It is also possible to deliver these beams with lower energies for treatment. Despite having ranges shorter than the human body, ions heavier than carbon, like oxygen and neon, could still be used for treatment at adjustable energies up to the full linac energy.

As for FLASH, ACCIL's capability is comparable to other existing proton and ion machines [20]. For example, for a proton beam of 230 MeV, losing about half of its energy in the last 10 cm, the energy deposited at $10^{10}$ p/s is $\sim$ 0.16 J/s. The corresponding dose delivered to a spot size of $\sim$ 5x5 mm2 (2.5 cm3 beam stopping volume) is 64 Gy/s, which is well above the FLASH dose requirement of 40 Gy/s. For a carbon ion beam of 430 MeV/u, losing about half of its energy in the last 10 cm, the energy deposited at $10^9$ p/s is $\sim$ 0.38 J/s, and the corresponding dose delivered to the same stopping volume is 152 Gy/s, which exceeds the FLASH dose requirement and calls for a larger beam spot size. However, to satisfy all cases, for all tumor sizes and beam energies, we would need at least 10 times more particles per second ($10^{11}$ protons/s and $10^{10}$ carbon ions/s), which is feasible with the ACCIL linac design. In addition, higher repetition rates may be required for faster beam scanning and more flexibility in beam delivery.

### 3.1.2 Enabling technology: Low-velocity high-gradient accelerating structure development

ACCIL requires the development of high-gradient structures (~ 50 MV/m) for ion acceleration with a relative velocity β in the 0.3-0.8 range. This makes the accelerating cells much more compact than β~1 cells built for electrons, especially at the lowest β. A shorter and more compact cell increases the rate of electric breakdowns and makes dissipating the power required for operation at such high gradients challenging. Research and development (R&D) in this field is being pursued at CERN [21], other European institutions, and more recently in the US by Radiabeam and Argonne [22]. In this collaboration, we have developed a β ~ 0.3 traveling-wave S-band structure (NHS) and demonstrated the 50 MV/m accelerating gradient required for ACCIL [23], see Figure 4. This special cavity design for the lowest velocity ions is what distinguishes ACCIL and makes it more compact than other linacs. It allows the transition to high-gradient acceleration to take place at 45 MeV/u, which is much lower than the 70 MeV/u for other linacs. At Argonne, we have also designed and prototyped a cold model of a β ~ 0.4 annular coupled structure (ACS) [24] as the next accelerating cavity for ACCIL following the NHS, see Figure 5,

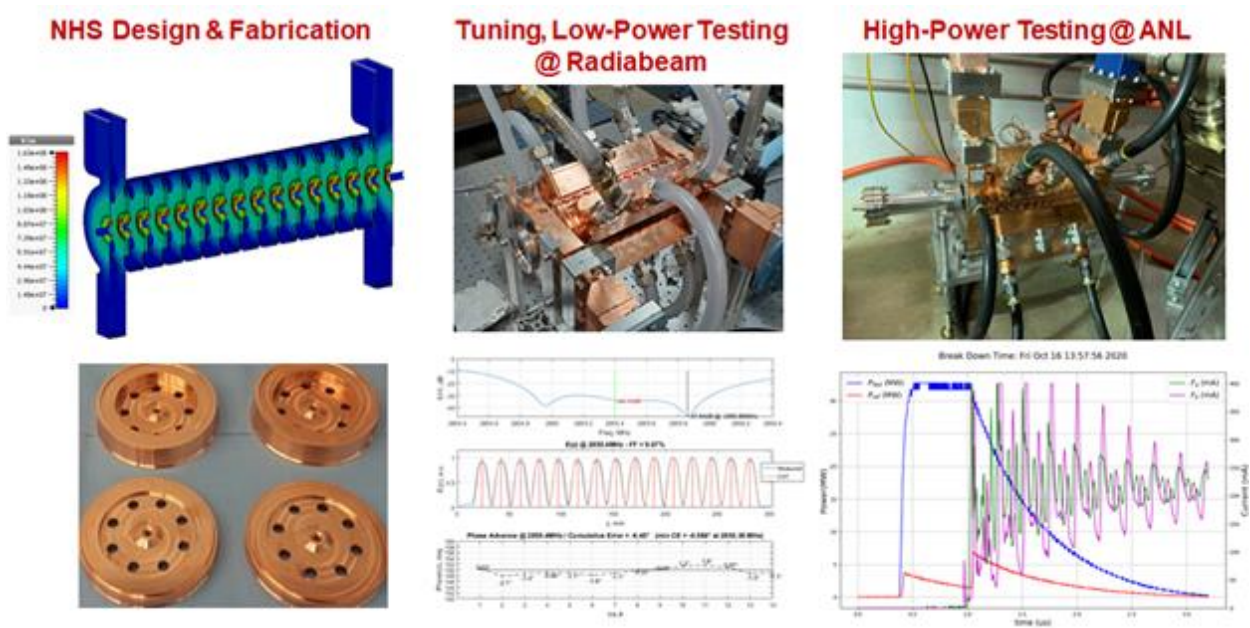

**Figure 4.** Design, fabrication, and high-power testing of the negative-harmonic traveling-wave structure (NHS) developed by Radiabeam in collaboration with Argonne.

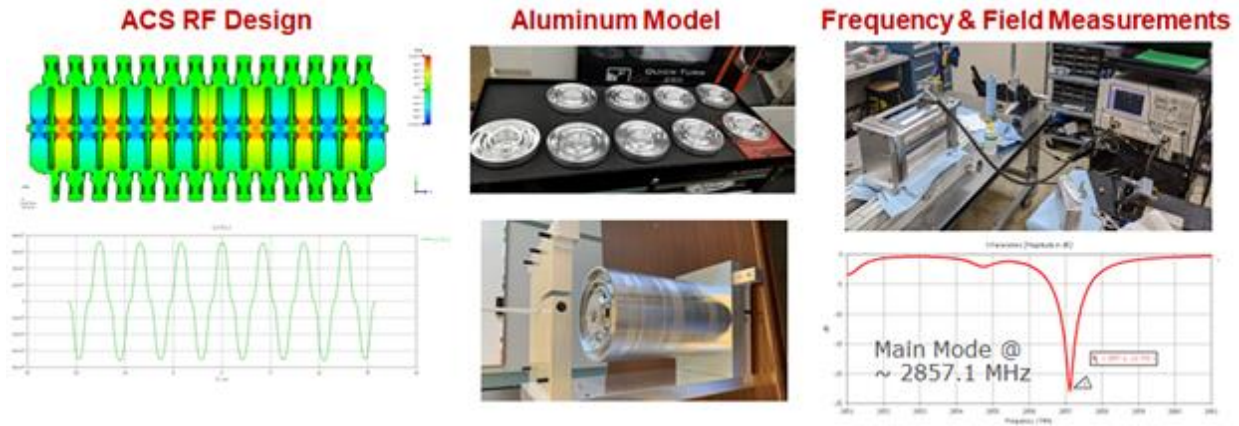

**Figure 5.** Design, cold model, and measurements of the annular-coupled structure (ACS) developed by Argonne in collaboration with Radiabeam.

### 3.1.4   Future developments - Path forward

In addition to the general development of high-gradient accelerator structures for low-velocity

ions, we identify a few areas of R&D of special importance for FLASH-RT with ion beams:

- Investigating and pushing the beam current limit of compact ion linacs

- Increasing the repetition rate of high-gradient structures

- Developing RF sources capable of delivering the required high pulsed power

More importantly and to enable this technology, establishing a linac-based advanced ion therapy research center in one of the National Labs would be a significant step forward and would allow the following:

- Cancer therapy and radiobiology research with all ion beams up to neon

- Radiography and tomography with ions lighter than carbon: proton, helium, …

- Real-time MRI guidance during beam delivery, significantly enhancing the outcome of ion beam therapy

- PET imaging using positron emitters (C-11, N-13, O-15, …) produced in the tumor for dose verification

- FLASH ion therapy (FLASH IT) and other novel approaches

We mention, in particular, an ACCIL-type linac that could be installed at the existing IPNS site at Argonne National Lab. with the required infrastructure [25], which represents a significant cost saving compared to a greenfield installation. Following the development and commissioning phases, an initial research program including cellular radiobiology and animal therapy could be conducted prior to human therapy and early clinical trials to prepare for FDA approval.

## 3.2 Fixed field gradient accelerators for FLASH-RT

### 3.2.1 Scaling fixed field gradient accelerator

Fixed Field Gradient Accelerators (FFGA), previously called Fixed-Field Alternating Gradient (FFAG) accelerators, are synchro-cyclotron style accelerators based on cycled radio-frequency acceleration. Similar to synchro-cyclotrons, FFAGs generally operate at high repetition rates, e.g., the new superconducting synchro-cyclotron (S2C2) from IBA operates at 1 kHz [26]. Compared to synchro-cyclotrons, a crucial difference of FFGAs is their strong focusing optics (no different from the optical principle of present days' pulsed synchrotrons) which results in much smaller beam size, as well as efficient handling of space charge defocusing effects, a concern when aiming at high charge bunches.

FFGA proton accelerators (aka "scaling FFAG") were developed at KEK in the late 1990s, with a proof-of-principle 500 MeV ring in 1999 [27] and a full-scale 150 MeV ring that provided the first beam in 2005 [28]. Two such rings are in operation in Japan, at Kyushu University (providing beam for condensed matter research) and at the Kyoto University research reactor (providing beam for KUCA, an ADS-Reactor Critical Assembly) [28]. These rings have demonstrated 100 Hz capability based on a single RF system; however, they may even do better as their lattice lends itself to multiple RF systems (in the manner of a folded linac).

Kyoto and Kyushu rings use so-called radial optics, an alternation of positive and negative bend radial sector strong focusing dipoles; compact rings, in addition, are obtained using spiral sector optics. This has been demonstrated by the RACCAM (Recherche en ACCelerateurs pour Applications Medicales) project [29], which has produced a design with a multiple-extraction ring (Figure 6), RACCAM constructed, as a proof-of-principle, a strong-focusing spiral sector FFGA dipole [30] and validated the design with 3D magnetic field measurements which proved that the expected performance was reached [31].

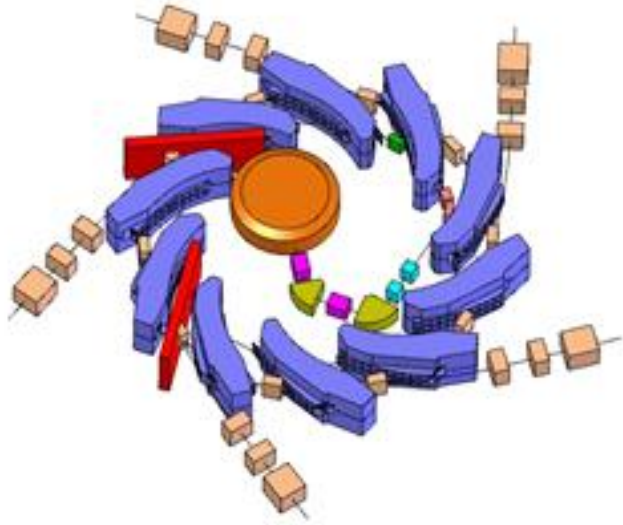

**Figure 6.** RACCAM multiple-extraction proton therapy FFGA ring [31].

The RACCAM spiral sector ring design allows >5 Gy$^x$liter/min dose delivery, based on bunch-to-voxel delivery. The reference volume (1 liter) is centered at 10 cm depth and comprised of 20$^x$20$^x$20, 5$^x$5$^x$5 mm$^3$ voxels [29]. This delivery mode requires a total of $10^{11}$ protons and for a uniform spread out Bragg peak $10^9$ protons per bunch in the most distal layer. Thus, the average dose rate is 5 Gy$^x$l/60 s ~ 0.1 Gy/sec over a liter.

The distal layer voxels require $10^9$ protons each (by contrast, upstream layers which take the dE/dx of more distal ones require decreasing bunch charge with decreasing depth), thus each $10^9$ proton bunch delivers 5 J/8000 ~7e-4 J; bunch length at extraction is~0.1 µs (about half a period of 5 MHz RF), this represents an instantaneous dose of ~7e-4/1e-7=4$^x$10$^3$ J/s. Thus, regarding

FLASH requirements, the average dose rate is by a factor ~200 low, and the instantaneous dose rate by about a factor of 40. Increasing the dose rate could be split between repetition rate, pushing beyond 100 Hz, bunch charge, pushing beyond $10^9$ protons per bunch (ppb), and decreasing the irradiated volume. In particular, with the assumption of 100 Hz repetition rate, a required <200 ms irradiation time imposes <40 voxels, 5 ml; current-wise this is $<I> = 10^9$ ppb$^x$e$^x$100 Hz=16 nA. An increase of the repetition rate by a factor of 10, to 1 kHz, would allow irradiation of a 50 ml volume irradiation in 200 ms; it would also bring the necessary bunch charge for proper average and instantaneous dose rates to $4^x10^9$ protons; current-wise, this is ~1 μA.

It can be seen from what precedes that increasing the average current (via repetition rate and/or bunch charge) and instantaneous current (via bunch charge) towards FLASH requirements is challenging. An intermediate option is to consider a smaller working volume, 10s of ml, for instance, which would relax on repetition rate and bunch charge constraints.

### 3.2.2   Nonscaling FFGA for FLASH

The constraints imposed by the field scaling law are relaxed in the nonscaling variant of the FFGA. Advanced codes and optimizers have been used to stabilize the machine tune consistent with isochronous orbits as in an iso-cyclotron. Isochronous orbits permit CW, high intensity beam, and the strong-focusing gradients allow long straight sections like a synchrotron. These straights can be used for high-gradient acceleration and low-loss, variable energy extraction using large-aperture bump magnets. The system design applied to therapy is described below.

*Overview*

A compact 250 MeV/nucleon, fixed-magnetic field turnkey machine has been designed in a racetrack format with variable energy continuous-output beam without a degrader, and with low-loss operation. The design is isochronous and produces continuous beam for ion species with a charge to mass ratio of ½ ($H_2^+$, $D^+$, $He^{2+}$, $Li^{3+}$, $B^{5+}$, $C^{6+}$, $N^{7+}$, $O^{8+}$, $Ne^{10+}$) and is therefore capable of accelerating all ion species to therapeutic energies. An outer ring can be added to further the energy reach of the ions to the full 430 MeV/nucleon.

Accelerating ions with an approximately constant charge to mass ratio has the advantage of equal beam transmission independent of ion species. Identical operation and extraction are maintained for all the therapeutic ions, including protons ($H_2^+$), implying turnkey operation even when switching between ions. Further, the possibility of accelerating and extracting multi-ion composite beams from a mix, or cocktail, of injected ion species (with an effectively equivalent charge to mass ratio) is ground-breaking technology. This approach also provides rapid switching between ion species, a capability based on <1% rapid adjustments in the RF frequency.

The complete system involves the injector, a preaccelerator, and the therapy ring – similar to cascade synchrotron systems. To allow variable energy extraction in a long straight section, the energy range must be restricted in order to extract inner, low-energy orbits using extraction magnets with feasible strengths and, in particular, apertures. The preaccelerator extraction energy not only facilitates variable-energy extraction in the higher energy ring but can also support an independent beamline for eye treatments and R&D.

*The High-Energy Therapeutic Ring*

The high-energy ring will be operated from 100 MeV/nucleon up to 250/330 MeV/nucleon to support ion therapy and particle imaging. This energy represents a 15 cm range for carbon and the energy required for light-ion imaging for pelvic or abdominal scans. The design of the therapeutic ring is a racetrack with opposing 5 m straight sections for RF and injection/extraction (2-fold periodicity). The ring also incorporates a 2m short straight in the center of each arc for vacuum and diagnostics. Figure 7 shows the layout of the ring and relative size compared to the Heidelberg Ion Therapy facility, which is a slow-cycling synchrotron capable of 430 MeV/nucleon. Particle tracking has been performed and has a large dynamic aperture with an acceptance>1000 mm-mr (normalized). Arcs can be either SC or normal conducting. SC extends the energy reach of the extraction system due to a reduced aperture and smaller distance between circulating orbits. However, the orbit separation needs to be studied and optimized for efficient and clean extraction of different energy orbits and the required acceleration gradient. An acceleration of ~2MeV/turn per nucleon appears to be a maximum step size requirement for longitudinal scanning. The isochronous level of performance in machine design shows less than a percent variation in TOF over an acceleration range from 70 MeV/nucleon to 250 MeV/nucleon.

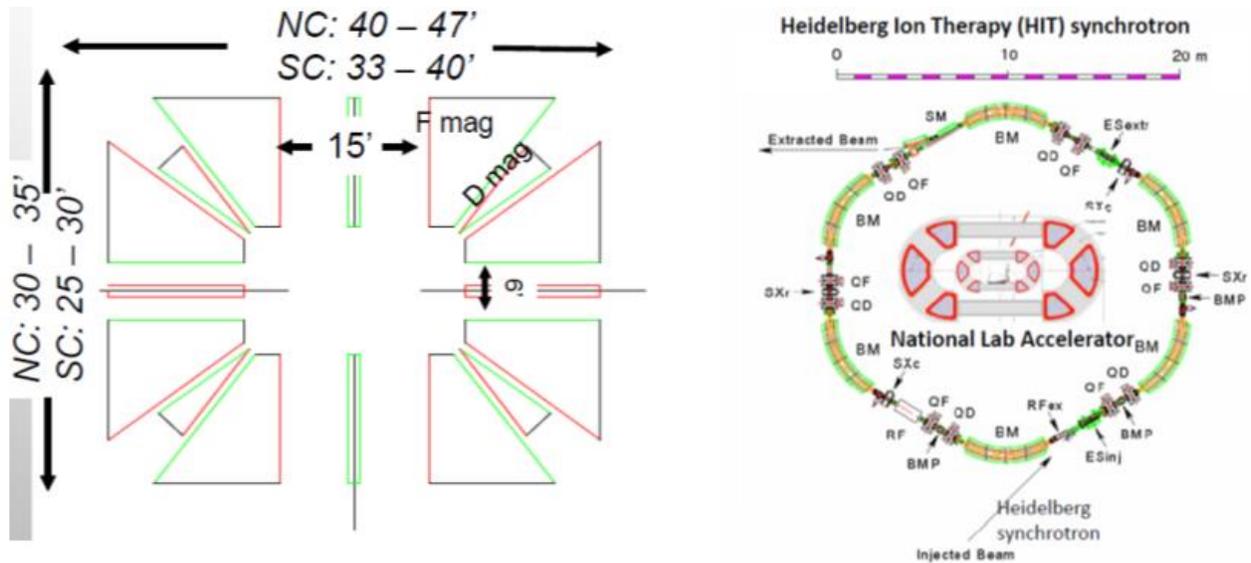

**Figure 7.** Outer dimensions for a variable energy 330 MeV/nucleon therapeutic ring (left) and a dual ring system 430 MeV/nucleon (right) compared with the Heidelberg Ion Therapy facility. On the right, the FLASH-capable, CW and variable-energy 430-MeV/nucleon ion accelerator nested system is compared to equivalent-energy, low-duty cycle Heidelberg Ion therapy Synchrotron. Inner ring racetrack is 250 MeV/nucleon and can provide independent beam delivery.

*Variable energy extraction*

Extraction is performed in one of the 5 m straight sections and shown in Figure 8. The magnets can be ramped for swept, variable energy longitudinal scanning or set at a flat-top for single energy beam delivery. The field direction is bipolar; field decreases and flips sign for maximum inner orbit extraction (blue to red lines). Extraction magnetic fields are limited to ~2.5T for a ramped system.

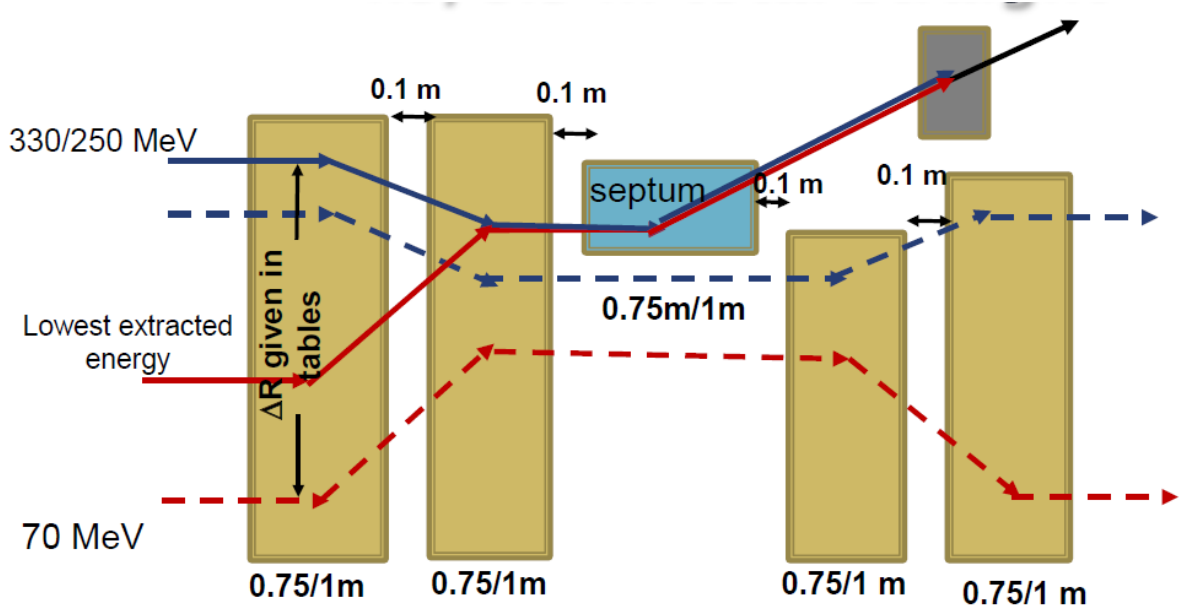

**Figure 8.** Layout of the ramped, bipolar magnet extraction system that selects the orbit and energy for extraction through a septum (left). Inner, lower-energy orbits are returned to their respective closed orbits for continued acceleration.

*Source, LEBT, RFQ, and Injector*

The ion source and beam capture system are comprised of an Electron-Cyclotron-Resonance (ECR) source coupled to a Radio-Frequency Quadrupole linac (RFQ) through a conventional Low-Energy Beam Transport (LEBT) section, as shown in Figure 9. The ion ECR source has one of the highest ionization efficiencies for gaseous elements. An RFQ linac, which uses electrical RF focusing, can capture, auto-bunch, and efficiently accelerate DC (constant current) ion beams directly from the source, achieving energies of several MeV, efficiently replacing complex and lengthy pre-injector elements. The LEBT will consist of an Einzel and solenoid lens system (beam chopping may be required to match RF frequencies between RFQ and injector). In addition, a mixture of ion species can be ionized in ECRs to produce a mixed or 'cocktail' of ion beams which potentially can be accelerated in the proposed accelerator system – including all therapeutic ions plus protons in the form of $H_2^+$ – thus combining imaging and therapeutic beams for real-time dosimetry.

An advanced, small-footprint, heavy-ion injector iso-cyclotron has been developed for the injector. This novel normal conducting, separated-sector injector has an optimized strong-focusing field gradient designed to efficiently accelerate light ions with a charge-to-mass near ½ (namely, protons in the form of $H_2^+$, $D^+$, $He^{2+}$, $Li^{3+}$, $B^{5+}$, $C^{6+}$, $N^{7+}$, $O^{8+}$, $Ne^{10+}$, $S^{16+}$ and $Ca^{20+}$) scalable up to 70 - 100 MeV/u [32,33]. Dual, high-gradient, 0.2 MV cavities with a tuning range of ±1% in frequency can accelerate any ion species with this charge to mass ratio on the 8th harmonic (~45 MHz) with large turn-to-turn, almost centimeter-level separation; an enabling compact and low-loss extraction technology that eliminates the charge-changing foils (used for injection in ion synchrotrons and extraction in H⁻ cyclotrons). Low (percent level or less) extraction losses are projected to be compared with the 20% - 60% (or even higher) losses of proton-therapy CW cyclotrons at extraction.

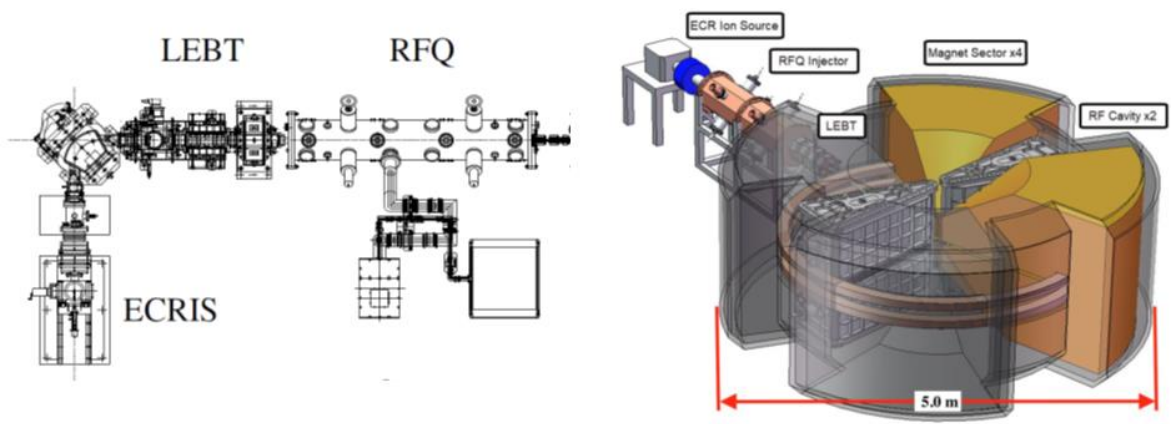

**Figure 9.** HIMAC, Japan ion source and RFQ (left), which serves as the concept for the upstream pre-acceleration system for the injector accelerator (right)A CAD model of the full conceptual design of the 20 MeV/u injector system (scalable to 70-100 MeV/nucleon) is shown on the right. From left to right, the ECR ion source, focusing solenoid, RFQ, beam focusing quadrupoles, and the cyclotron are shown with a transparent outer shielding for clarity.

*Summary*

A complete CW, variable-energy ion therapy concept has been developed with the preaccelerator stage design advanced in terms of engineering. Since FLASH intensities have been achieved and the effect observed using proton iso-cyclotrons, this ion therapy complex is FLASH capable. Further, it has the advantage of supporting essential R&D beyond the shoot-through beams currently available for hadron R&D. In addition to providing intense ion beams, FLASH radiotherapy studies can be extended to incorporate the Bragg peak range dependencies and dosimetry into a broader research initiative. The injector and higher-energy therapy accelerator being developed for the National Particle Beam Therapy Center (Waco, TX) will provide the range of ions intensities with different LET and relative biological effectiveness (RBE) requested by the medical community in a CW beam format without significant operational modifications and overhead; i.e., a turnkey system. This system represents a significant advance in ion therapy clinical application and plays a critical role in the development of FLASH-RT.

## 3.3   FLASH studies with laser-driven particle sources

### 3.3.1   Status of laser-driven particle sources

Novel laser-driven particle sources are receiving increasing attention due to their potential of providing particle beams for applications on a relatively small footprint and at a potentially lower cost than radiofrequency (RF) driven accelerators [34,35]. Efficient laser-particle acceleration has become feasible with the advent of ultra-short pulse high-power lasers enabled by chirped pulse amplification [36], a technology that was awarded the Nobel Prize in Physics in 2018, which yielded peak laser powers exceeding 1 petawatt (PW) [37]. The most prominent acceleration schemes are laser wakefield acceleration (LWFA) of electrons [38] and target normal sheath acceleration (TNSA) of protons and ions [39]. LWFA is conducted with gas targets that are quickly ionized by the leading laser pulse edge, followed by the formation of collective plasma oscillations in the wake of the pulse as it propagates through the transparent plasma. Free electrons are accelerated up to several GeV energies in dynamic electric fields associated with the resulting plasma wave [40,41]. If optimized, monoenergetic electron bunches with nC charge can be generated [42]. At the time of writing, Lawrence Berkeley National Laboratory held the LWFA electron energy record, with 8 GeV and a few pC charge [43]. LWFA sources can be used to drive compact light sources from the high-field THz [44] over the high-brightness X-ray [45] to the gamma ray range [46].

Target normal sheath acceleration (TNSA) is generally pursued with solid targets, most commonly in the form of a few μm thick metal or plastic foils. Equally ionized by the lead edge of the laser pulse, the laser peak intensity interacts with free electrons in a preformed plasma layer at the target surface. Electrons gain energy in the laser field, circulate through the target bulk and expand beyond the predominantly fixed ion distribution at the target surfaces. The resulting quasi-static charge separation fields are in the order of TV/m and lead to the acceleration of protons and ions to >10 MeV energies, emitted along the target normal with a beam divergence of roughly ± 20° [47,48]. The generated proton beams are of high flux (up to $10^{13}$ particles per pulse [49]) and feature broad exponential energy spectra up to the characteristic cutoff energy, approaching 100 MeV [50-53].

At current PW laser pulse repetition rates of at most 1-10 Hz, directing laser-driven (LD) particle beams to biological samples results in a moderate *mean* dose rate. However, due to their generation mechanism, resulting in ultra-short particle pulse lengths of less than a picosecond at the source, LD particle beams naturally feature ultra-high *instantaneous* dose rates (IDR), exceeding $10^9$ Gy/s. This IDR is several orders of magnitude higher than dose rates typically delivered with RF driven accelerator technology [54].

### 3.3.2   Using laser-driven particle sources for preclinical radiobiological studies of the FLASH effect

LD particle sources may soon become adequate complements to RF driven accelerators for basic radiobiological research of the FLASH effect [55]. Access to conventional experimental and medical machines has been rather limited for this type of research [56], while the steady increase in available compact LD particle sources has already started to open up new experimental options for systematic radiobiological studies.

The majority of radiobiological studies with LD particle sources has been conducted in view of potential future applications in radiotherapy, in particular with protons and heavier ions. As such, an appreciable number of *in vitro* studies have been conducted to investigate the radiobiological effectiveness of LD protons [57-68]. Fewer radiobiological studies were so far conducted with LD electrons [69-74]. The only *in vivo* study found no difference in tumor growth delay comparing LD electrons and RF accelerated electrons [71]. The proposal of using very high energy electrons (VHEE) with energies in the range of 150 - 250 MeV for radiotherapy [75] has sparked renewed interest in the dosimetric properties and the potential for new radiotherapy protocols using compact LWFA electron sources [76-78]. So far, no differential sparing effect of normal tissue was reported from radiobiological studies with LWFA electrons.

While the dose rate was not always specified in publications, it can be assumed that samples were irradiated at ultra-high IDR due to the LD particle acceleration mechanisms, as mentioned above. So far, the majority of radiobiological studies with LD particle beams were conducted using *in vitro* cell cultures and at atmospheric ~ 20% oxygen levels.

Magnetic transport beamlines have been implemented at a few laser facilities to transport LD protons to a dedicated sample site and apply a three-dimensional dose profile for *in vivo* studies with small animals, for which tumor models have been developed [79-82].

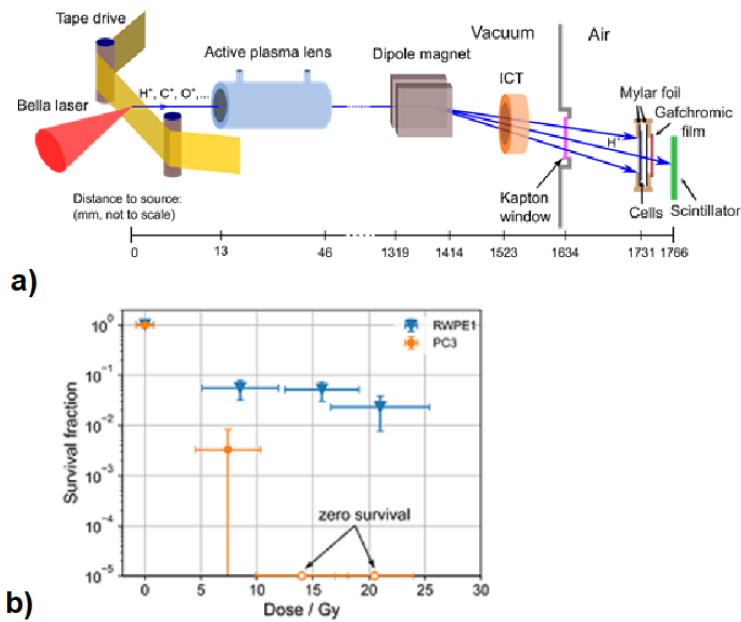

**Figure 10.** Figure adapted from Ref. [68]. a) Schematic depiction of the laser-driven proton beamline at the BELLA PW laser with tape drive target, active plasma lens, dipole magnet, integrating current transformer (ICT), cell sample, gafchromic film, and scintillator. b) Cell survival fraction of human prostate cancer cells (PC3) and normal human prostate cells (RWPE1) after irradiation with laser-driven protons. c) Proton beam parameters for cell sample irradiations at the BELLA PW

**Table 3.** Proton beam parameters for cell sample irradiations at the BELLA PW

| Dose per pulse | 1 Gy |
|---|---|
| Pulse length | 30 ns |
| Pulse repetition rate | 0.2 Hz |
| Instantaneous dose rate | $3 \times 10^7$ Gy/s |
| Mean dose rate | 0.2 Gy/s |

In a preliminary study at the 40 Joule BELLA petawatt laser proton beamline, it was demonstrated for the first time that LD protons delivered at ultra-high IDR can indeed induce the differential sparing of normal versus tumor cells *in vitro* for total doses ≥ 7 Gy [68]. In that study, normal and tumor prostate cells in 1 cm diameter custom cell cartridges were irradiated with LD protons of 2-8 MeV at an IDR of $10^7$ Gy/s. After acceleration from a tape drive target, the proton bunch was transported with a compact active plasma lens beamline [83] to the cell sample site located outside the vacuum chamber (Figure 10 a). An integrating current transformer (ICT) was implemented for online beam charge measurements. The dosimetry was performed with calibrated radiochromic films attached to every cell sample. With 1 Gy applied per laser shot, total dose values up to > 30 Gy were accumulated by operating the LD proton beamline at 0.2 Hz. A significant sparing of normal prostate cells compared to prostate tumor cells was observed after irradiation with LD protons (Figure 10 b). The main proton beam parameters for this study are summarized in Figure 10 c. Reference irradiations with X-rays at clinical dose rates did not show a similarly differential radiosensitivity. It should be pointed out that the generation of LD proton beams in the energy range sufficient for this type of study does not require a PW laser system but was demonstrated in numerous experiments at 100 TW-class laser systems [52,53].

### 3.3.3   Potential of laser-driven particle sources for FLASH Radiation Therapy

Preliminary *in vitro* experiments with LD ion sources have shown promise for FLASH radiotherapy [68]. However, stringent requirements concerning combined key beam parameters like proton energy (up to 250 MeV), numbers of protons per bunch ($10^9$), stability and control of energy and proton numbers from shot to shot (< few percent), and repetition rate (> 10 Hz) are yet to be experimentally demonstrated [84].

Currently, the primary challenge for the field of laser-ion acceleration is reaching clinically relevant particle energies. So far, peak LD proton energies achieved in experiments are approaching 100 MeV [50], which is well below energies necessary for clinically relevant penetration depths of >30 cm in humans [85]. Ongoing efforts to develop a high-repetition rate, several PW-class lasers can theoretically overcome this challenge when combined with improved gantry designs and treatment planning strategies specific to LD particle sources [86,87]. Currently, no unified reference dosimetry protocol exists for LD particle beams, which are unique in their ultra-high IDR and, in the case of ions, broad energy spectra. However, innovative dosimetry methods for radiobiological studies with LD proton sources have been developed that use online, minimally invasive, relative dose detectors, e.g., thin transmission ionization chambers, corrected for recombination effects [88] and cross-referenced with independent absolute dosimetry methods like radiochromic films [89] or Faraday cups [90]. These have enabled *in situ* dose-controlled LD proton irradiations of biological cell samples at a relative dose uncertainty below 10% [91].

Advances in laser technology are expected to deliver higher LD proton and ion energies because experiments and simulations have shown a consistent increase of maximum particle energies with laser pulse energy, power, or intensity [52,53]. At the same time, theory and simulations predict higher proton and ion energies when harnessing advanced acceleration regimes, including, for example, radiation pressure acceleration [92], magnetic vortex acceleration [93], and shock acceleration [94].

Given that the aforementioned ion source requirements are met, designing a hypothetical compact LD FLASH radiotherapy machine requires careful consideration of not only the laser particle source but also the treatment beam delivery system that needs to reliably shape a 6-dimensional dose profile matching the tumor profile of which first designs exist [86]. As such, even after optimizing LD particle sources for footprint, which on the laser side may come naturally with advances in laser technology, it remains to be seen whether they can compete in compactness and cost with emerging conventional proton therapy machines, e.g., compact solutions by Mevion, IBA, Hitachi, and others, which have seen significant developments in recent years to reduce their footprint and cost [84]. However, these machines are typically not able to deliver comparable IDR as LD proton sources. So far, no such compact machines exist for heavier ions, which are automatically accelerated alongside protons in LD ion accelerators.

With the current interest in using VHEE for radiotherapy, LWFA may well offer the most promising method for compact and affordable VHEE medical machines that can operate in the ultra-high IDR regime [35]. While the necessary electron energies are readily generated in a well-controlled laboratory setting, the long-term source stability and reproducibility require further improvement. Moreover, current limitations to the achievable mean dose rate due to lower repetition rates compared to RF driven accelerators, in combination with a lower, less localized energy deposition compared to ions, need to be addressed. Current efforts towards high average power, Joule-class kHz laser systems may provide solutions for some of these issues [95].

### 3.3.4 Summary

To summarize, current LD particle source parameters are well below the requirements for their use as an alternative medical FLASH radiotherapy modality. However, their comparatively low-cost and compact nature has earned LD particle sources increasing attention and the differential normal tissue sparing *in vitro* under LD proton irradiation was recently demonstrated [68]. Therefore, LD particle sources could soon complement conventional accelerators to increase and democratize access to particle sources for preclinical radiobiological research. This real-world application can serve as a stepping stone to further advance LD particle sources to the necessary capabilities to provide particle beams for FLASH radiotherapy.

### 3.4. High peak current linear induction accelerator (LIA) for FLASH-RT[1]

### 3.4.1 Introduction

Meeting the requirements for reproducible FLASH effects of $>1.8 \times 10^5$ Gy-s$^{-1}$ instantaneous dose rates with an overall irradiation time <200 ms (>40 Gy-s$^{-1}$ average dose rate for healthy tissue sparing) using deep penetrating MV bremsstrahlung (i.e., X-rays) requires tens of ampere pulsed electron beams at high pulse repetition frequency (PRF) [2,96]. These dose rate criteria must not only be met in the core of the irradiated volume but in the whole of the volume as well (i.e., periphery and exit edge) [12].

---

[1] Patents Pending

The pulsed power based linear induction accelerator (LIA) using a multilayered bremsstrahlung conversion target meets these demanding requirements. Complementary irradiation sources from the same accelerator structure ensure that the whole of the irradiated volume is above the FLASH threshold. The LIA acceleration technique stores energy during the interpulse time and then discharges it in a short, 10s of nanosecond pulse to achieve extremely high instantaneous dose rates. This method is a direct acceleration technique using *induction* and does not require RF or microwave generation as an intermediate step. When operated at a clinician-specified kilohertz PRF, equivalent dose rates exceeding the healthy tissue sparing thresholds can be easily achieved, and the concentration of specific radicals, thought to play a role in the FLASH process, can be manipulated (see for instance [97]). Using active control of the individual pulses ensures safe dosing. While not widely known to the medical community, existing systems have been used as 10s of MeV, kiloampere electron or ampere level hadron sources since the 1960s [98].

A key demonstrated capability of the technology is that the beam pipe can be made arbitrarily large without affecting the acceleration gradient [99,100]. This property enables independently controllable multibeam acceleration through a single accelerator structure for complimentary irradiation. While the gradient of older induction linacs is low (<1 MV/m for 50-70 ns pulses), modern approaches enable 5-10 MV/m gradients. Thus, a 16 MeV system with 4-8 or more individual beams with variable energy and pulse rate would fit in a clinic-sized vault of ~100 m$^3$.

Figure 11 depicts an artist's conception of a four beam system; added beams are easily implemented [101]. The accelerator cells can be seen beyond the patient and patient couch. Four separate electron beams are being accelerated away from the patient, captured, and then bent 180$^\circ$ with two 90$^\circ$ dipoles. Solenoid transport is then used to return the beam alongside the patient, and a third 90$^\circ$ dipole directs the beam to a target where it is converted to bremsstrahlung. Multi-leaf collimators (MLC) can be used in this region for conformal therapy

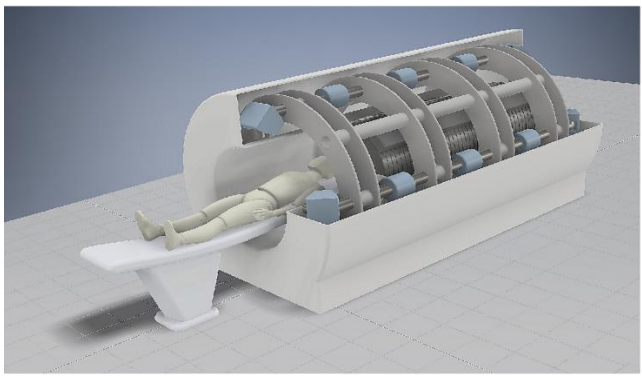

| Parameter | Induction Linear Accelerator |
|---|---|
| Electron Energy (MeV) | 16 |
| Total Beam Current (A) | 25 |
| Pulse Width (s) | 1.50E-08 |
| Pulse Repetition Frequency (Hz) | 10000 |
| Net Gradient (MeV/m) | 5.0 |
| Accelerator Length (m) | 3.20 |
| Inner Radius (cm) | 10.0 |
| Instantaneous Surface Dose Rate (Gy/s) | 6.60E+05 |
| Average Surface Dose Rate  (Gy/s) | 98.9 |
| Total Dose (Gy) | 19.8 |
| Time On (s) | 0.20 |

**Figure 11.** Concept FLASH-RT system using a linear induction accelerator (LIA) providing four or more lines-of-sight. LIA is on axis with the patient. Blue components are the magnetic focusing elements that direct the electron beam to the patient. The active accelerator is 3.2 m. With the returning drift section, the overall system length is 3.5 m less the patient couch [101].

### 3.4.2  Illustrative measurements from LIAs for FLASH-RT

Bremsstrahlung generated by an LIA provides broad area, deep penetrating, and high dose rate capability. This unique capability results from the elimination of resonant structures characteristic of the majority of acceleration techniques. Such structures are prone to pulse shortening beam instabilities and also degradation of the acceleration gradient when beam currents approach one ampere [102-104]. We present measurements on the FLASH X-ray (FXR) accelerator used to accelerate electrons to 17 MeV and briefly describe the Experimental Test Accelerator-II (ETA-II) with a nominal output energy of 6.5 MeV but at kilohertz PRF [105 -107].

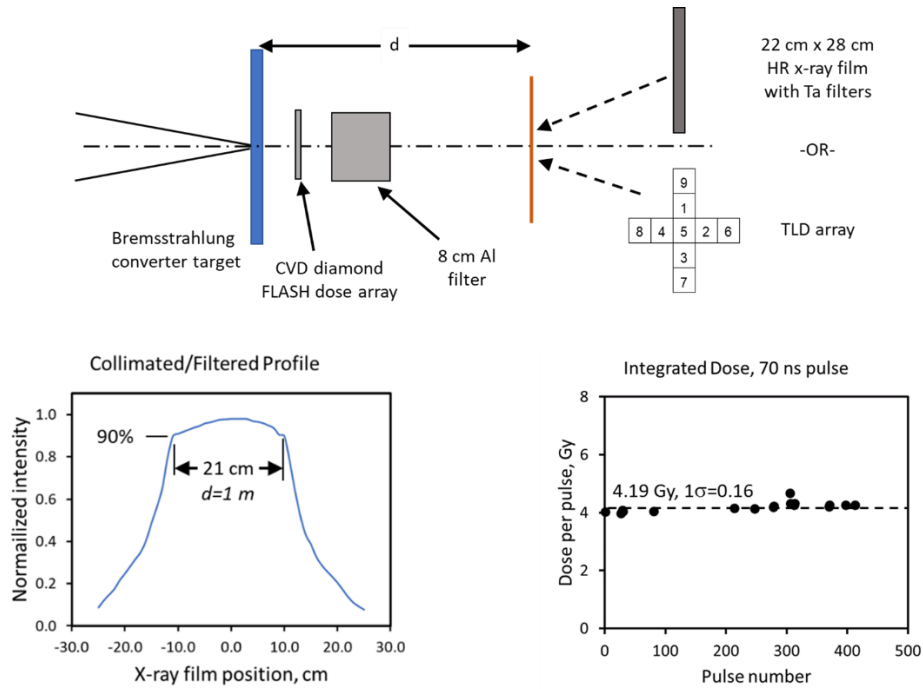

**Figure 12.** Measurement geometry, bremsstrahlung field flatness, and pulse-to-pulse repeatability.

The FXR geometry and measurements are shown in Figure 12. This particular FXR arrangement consisted of the bremsstrahlung converter target, a fast CVD diamond FLASH dose detector, an 8 cm thick low energy filter, and either thermoluminescent (TLD) or film detector at 1-2 m [107]. The 90-100% flat field is approximately 21 cm diameter at 1 m. This measurement corresponds to an approximately 350 cm$^2$ area demonstrating that FLASH levels can be maintained in the totality of an irradiated volume. We observe a *single shot, stable* dose of approximately 4.19 Gy with 1σ≈0.16 or a 3.9% variation. This value corresponds to an instantaneous dose rate of approximately 6 x 10$^7$ Gy-s$^{-1}$.

On ETA-II, a 5 kHz PRF has been demonstrated. This system also produced highly stable electron beams with less than 1% energy variation, millimeter spot size, and submillimeter spot motion. Initial use of the accelerator was in conjunction with a wiggler to generate electromagnetic energy at 2 GW and 140 GHz mm-wave energy for fusion research studies [108].

### 3.3.5   Meeting the criteria for FLASH-RT

To ensure that both the periphery and exit dose rate were above the healthy tissue sparing threshold to minimize toxicity, we performed calculations assuming a minimum of four separate sources placed symmetrically around a water phantom volume approximating the average human torso 16 cm radius [12,109]. Each source provides 25 Gy-s$^{-1}$ at 1 m using a total beam current of 25 A; details are provided elsewhere [101]. This configuration achieved 50% beyond the required average healthy tissue sparing dose rate, or about 60 Gy-s$^{-1}$ throughout most of the volume. By contrast, a single source at the same level achieved the healthy tissue sparing dose rate >40 Gy-s$^{-1}$ nearest the source, but about 25% of the volume is below that dose rate, potentially inducing toxicity. Noting that the total beam current for LIAs typically exceeds 1 kA, current can be used as a free parameter for increased dose rate.

The model assumed a single LIA to accelerate separate beamlets in an approximately 14 cm diameter beam pipe. While four beams are shown, eight or more can be easily implemented in an actual system. Beam transport is managed through the accelerator with solenoid coils and integrated steering similar to the FXR geometry [105]. The added steering capability allows generating oblique rays to allow a closer approximation of multibeam conformal therapy. Based on the model, at 16 MeV, the system would be approximately 3.2 m long (Figure 11), delivering a uniform average dose rate of 60 Gy-s$^{-1}$ at a beam current of 25 A at 10 kHz PRF.

**Acknowledgments** - This work was performed under the auspices of the U.S. Department of Energy by Lawrence Livermore National Laboratory under Contract DE-AC52-07NA27344. Work performed by Opcondys, Inc. was funded separately by the United States government under ARPA-E (Contract No. DE-AR0000907), the National Science Foundation (Grant No. 1519964), and the California Energy Commission CalSEED Program (Grant No. 17-01-03).

## 3.5  High-current electron linear accelerator for X-ray FLASH radiation therapy

### 3.5.1  X-ray FLASH-RT

An attractive tool for delivering FLASH-RT could be a FLASH-capable X-ray system. More than 80% of all radiotherapy is delivered with X-rays [110]. They are the most versatile form of radiation therapy and the most cost effective. Unfortunately, the physical process for generating X-rays is not very efficient [111]; therefore, a high-power accelerator is needed for X-ray FLASH-RT. Furthermore, one would still like to achieve as much conformality as possible. Conformality, combined with the healthy-tissue-sparing FLASH effect, promises to dramatically improve patient outcomes [112].

Considering the inevitable reduction in effective dose rate with intensity modulation and transmission through small apertures, a linac that can deliver 100 Gy in one second or faster is challenging but not impossible. Conventional 6 MV medical linacs produce a flattening filter free dose rate of around 0.2 Gy/s at one meter from the X-ray target – three orders of magnitude too low – however, they are on the low end of the spectrum of linac powers [113]. A typical medical linac has a beam power on the order of 1 kW. In comparison, industrial accelerators for sterilizing food and medical products can achieve beam powers of several hundred kW [114].

Another factor that allows for improvement in dose rate is increasing the beam energy. The conversion efficiency from electron beam power to X-ray power scales approximately with $E^3$, so a small increase in energy can make a big difference in X-ray intensity. The higher X-ray energy also allows greater penetration. However, there are two major downsides to higher photon energies: larger lateral penumbra (a measure of the fuzziness at the edge of the beam) and greater neutron production (which causes activation and unwanted dose to the patient and the environment). Photon energies up to 20 MV are commonly used in RT. We consider 10-18 MV to be optimal for achieving a high dose rate while limiting the negative factors.

One could also consider reducing the distance from the source to the target. However, this can only be done to a certain point without sacrificing useability. Achieving good conformality requires placing one or more collimators between the beam source and the patient. Along with pure physical limitations on fitting the equipment around the patient, this limits the source-to-surface distance (SSD) to 80 cm at the smallest.

### 3.5.2  New technology for X-ray FLASH-RT

RadiaBeam and UCLA are working on a solution for X-ray FLASH therapy that takes advantage of a single linac based on already-demonstrated technology and an innovative yet straightforward method for intensity modulation [115]. The major innovation of the proposed project is the development of the Rotational direct Aperture optimization with a Decoupled (ROAD) multi-leaf collimator (MLC) ring [116]. Intensity modulation has been the key driver in improved patient outcomes in RT over the past three decades, but there has been no solution to do this in the short time required by FLASH. With ROAD, the linac pulses are timed to align with a counter-rotating ring of 75 pre-shaped MLC apertures. As both the linac and MLC ring rotate in opposite directions at 60 rpm, 150 modulated beams are delivered in 1 s, each delivering up to 0.7 Gy to the tumor. ROAD can achieve physical dose conformality superior to state-of-the-art VMAT plans free from the MLC mechanical limitation, yet with the added benefit of the FLASH effect. Figure 13 shows a model of the proposed ROAD-FLASH system [13].

The linac (see Table 1 for a summary of parameters) consists of a 1.3 A, 140 kV electron gun, prebuncher, chopper, and two traveling wave linac sections powered by a commercially available 20 MW L-band klystron with 167 μs pulses at 150 Hz, to bring an 8.14 mA average current electron beam to 18 MeV. Assuming a dose conversion factor of 2,000 Gy/min/mA at 18 MeV, such a linac will be able to provide an uncollimated dose rate of 271 Gy/s at 1 m (8.14 mA $\times$ 2 · $10^3$/60 Gy/s/mA), which is equivalent to ~100 Gy/s collimated dose at 80 cm, assuming ~25% dose delivery efficiency. The beam is transported through a rotary vacuum joint into a rotating magnetic gantry that brings the beam to a rotating X-ray target directed at the patient.

**Table 3.** Parameters of ROAD high-current electron linear accelerator for X-ray FLASH therapy

| System | ROAD | Conventional [10] |
|---|---|---|
| Energy [MeV] | 18 | 6 |
| Pulse Length [µs] | 167 | 4 |
| Rep Rate [Hz] | 150 | 250 |
| Duty Cycle | 2.5% | 0.1% |
| Injected current [A] | 1.3 | 0.5 |
| Transmission | 25% | 25% |
| Peak current [A] | 0.325 | 0.125 |
| Average Current [mA] | 8.14 | 0.125 |
| Dose Rate Factor [Gy/min/mA at 1 m] | 2,000 | 120 |
| **dose rate, uncollimated, at 1 m [Gy/s]** | **271** | **0.25** |
| Dose Delivery Efficiency | 25% | 25% |
| **dose rate, collimated, at 80 cm SAD [Gy/s]** | **106.0** | **0.10** |

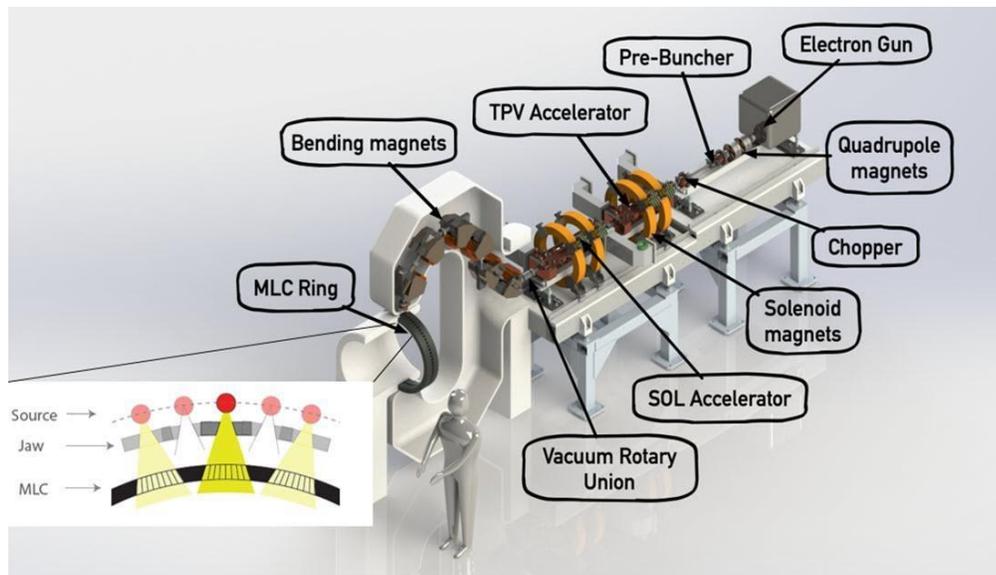

**Figure 13.** Rendering of the ROAD FLASH-RT system. A segment of the decoupled MLC ring is shown in the figure inset with three MLC modules. The linac is triggered to produce the beam when the target is aligned with the MLC to produce a VMAT-like treatment.

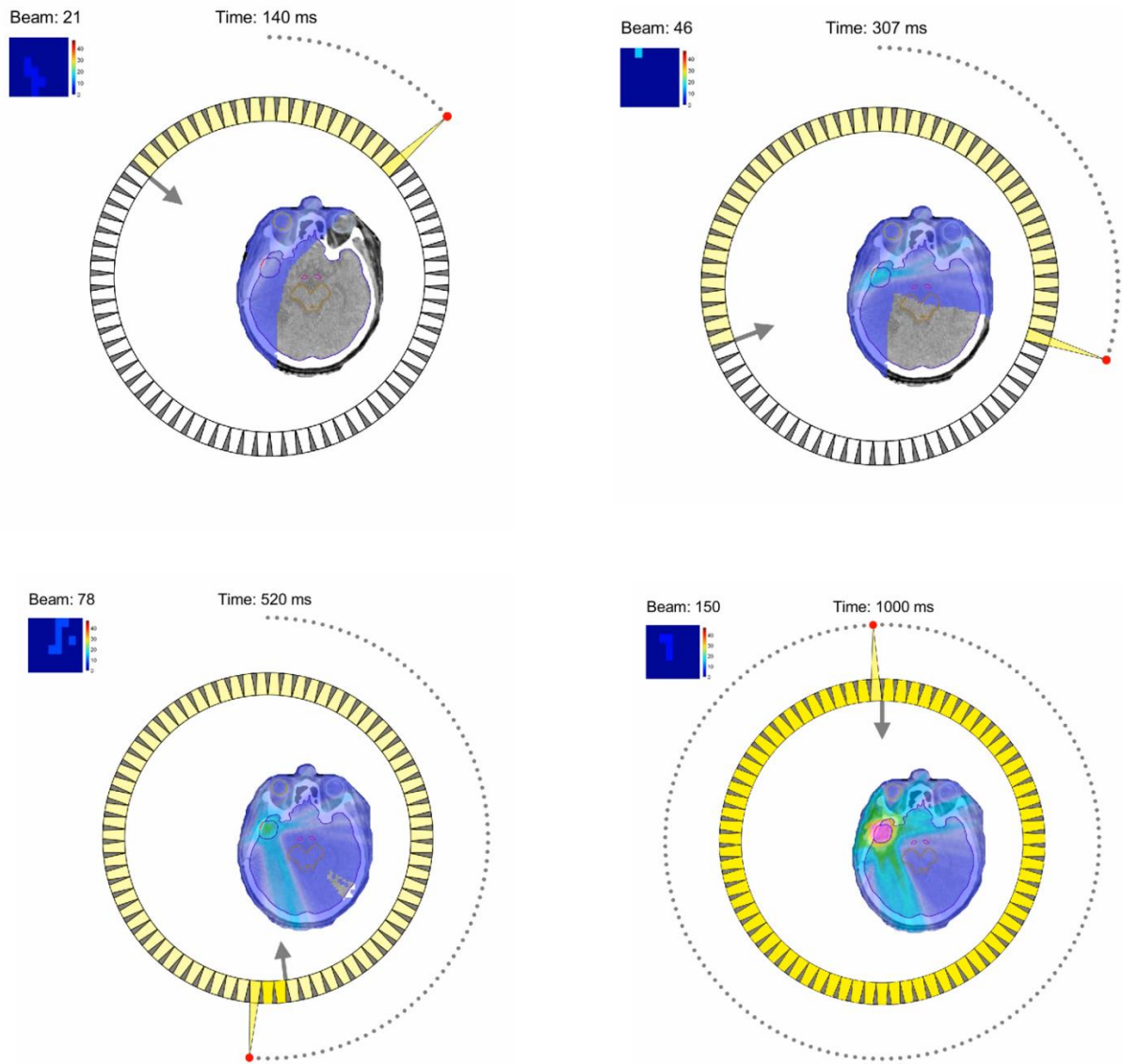

**Figure 14.** FLASH-RT delivery using ROAD. The figure shows four beams with the corresponding pre-shaped MLC in the upper left corner of respective beams. Cumulative radiation dose to a brain tumor patient is shown in the center.

Figure 14 shows the proposed delivery FLASH delivery using the ROAD method. There are a total of 75 MLC modules on a ring that is separate from the X-ray source. The X-ray pulses are triggered when the source is sequentially aligned with the MLC apertures. Counter-rotating the MLC ring allows more aperture shapes to be programmed for further improved physical dose conformity.

With an increasing understanding of the underlying FLASH mechanism, it is necessary to further quantify the FLASH effect at the treatment planning stage as part of the inverse optimization goal. The feasibility has been demonstrated for simultaneous dose and dose rate optimization (SDDRO) with protons [117] and X-ray ROAD FLASH [116]. In the latter, the hypothesized oxygen depletion effect was parameterized into the planning system to show a larger FLASH effect for normal tissue sparing with larger individual pulses. The information will guide the design of high output and FLASH-ready linear accelerators.

### 3.6  Accelerator-based technology developed at SLAC National Accelerator Laboratory and Stanford University

#### 3.6.1  X-ray FLASH-RT with the PHASER

The inherent inefficiency of producing therapeutic X-rays through bremsstrahlung radiation from an electron beam hitting a target contributes to the challenge of achieving FLASH dose rates with conventional photon radiotherapy. Another critical factor constraining the treatment time in a conventional radiotherapy device is the mechanical motion of the gantry. The PHASER program for Pluridirectional High-energy Agile Scanning Electronic Radiotherapy, led by SLAC National Accelerator Laboratory and Stanford Medical School, seeks to eliminate gantry motion while achieving the FLASH dose rate through a system of 16 linacs arrayed around the patient, as shown in Figure 15 [118]. This design enables multiple angles of approach, as needed for intensity modulated conformal radiation therapy. The PHASER program is intended to increase the therapeutic index of radiotherapy through highly-conformal image guided FLASH treatments and improve the accessibility of state-of-the-art FLASH capable medical equipment through the implementation of a compact and economical design.

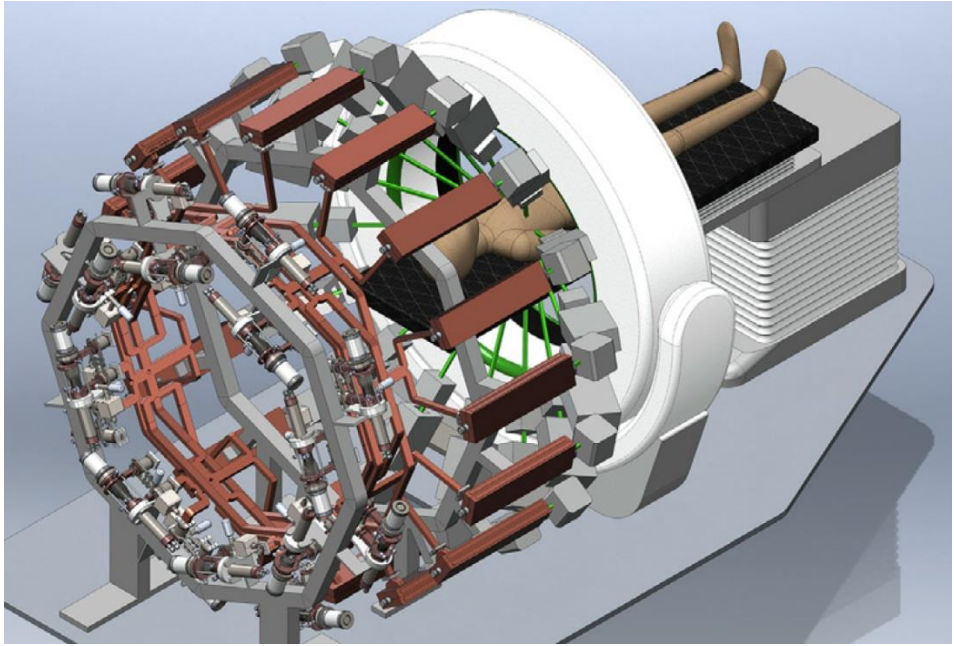

**Figure 15.** Conceptual diagram of the PHASER architecture with 16 linacs arrayed around the patient for highly conformal FLASH photon therapy.

PHASER seeks to achieve FLASH capability by producing a >400-fold increase in average beam current compared to conventional 10 MV photon therapy systems with a typical dose rate of 10 Gy/min. The linac structure will rely on a distributed coupling topology to improve the power efficiency, allowing the system to take advantage of a network of 16 compact "klystrinos," each producing a peak power of around 330 kW. The research and development from the initial PHASER program has laid the groundwork for extending the design concept to a Very High

Energy Electron (VHEE) therapy system, treating directly with ≥100 MeV electrons without x-ray conversion.

### 3.6.2   Very High Electron Energy (VHEE) FLASH-RT

Very High Energy Electron (VHEE) radiotherapy is a key area of opportunity to apply technology developed for the particle physics community to the new field of FLASH-RT. Direct use of an electron beam for radiotherapy provides one of the most readily scalable paths to achieve FLASH dose rates, as apparent in the existing body of experimental evidence for FLASH that has been predominantly performed with direct electron beams [1,5,119,120]. Existing facilities have been modified for FLASH radiotherapy experiments [121-123], and commercial systems like the Oriatron eRT6 from PMB-Alcen [124], Kinetron from CGRMeV [125], Novac7 from Sardina [126], and the Mobetron from IntraOp [127] have been employed for FLASH capability at sub-10 MeV electron energies. While clinical and pre-clinical commercial devices proceed with development to provide ultra-high-dose rate direct electron therapy [128,130], treatment of superficial tumors in human patients has already begun [2].

The key technological advance where active research is needed concerns the development of medical accelerators that can reach the high beam energies required to treat deep-seated tumors throughout the body. The electron beam energy determines the penetration depth, with a 100 MeV beam reaching a depth of about 40 cm, sufficient to cover almost all deep-seated tumors [131,132]. VHEE therapy has yet to be realized in a clinical setting because existing equipment lacks the capability to reach these beam energies. The maximum energy obtained for direct electron FLASH-RT thus far has been only 20 MeV, using a modified Varian Clinac [119,120]. The size and power requirements to simply extend this structure to produce 100 MeV beams would be prohibitive for clinical use. The high beam energies used for VHEE treatments will also impose additional shielding requirements on the treatment facility. Treatment planning studies using Monte Carlo simulations indicate that a dose rate of approximately $2 \times 10^4$ Gy/s can be achieved per milliamp of average beam current over a 10 cm by 10 cm field size.

Current RF-driven linear accelerator research programs aimed at meeting the demand for VHEE capability have primarily focused on advanced, normal-conducting, high-gradient accelerator RF technology. A CHUV-CERN collaboration to build the DEFT (Deep Electron FLASH Therapy) facility plans to combine an X-band linac design developed through CLIC research with an S-band photoinjector [133]. Designs for a VHEE system at the PRAE accelerator utilize an S-band linac in addition to an S-band photoinjector, prioritizing linac performance reliability [134]. Implementation of a photoinjector with a medical linac has yet to be demonstrated, but the strategy of utilizing a photoinjector has already been proposed as an opportunity to harness the speed and flexibility of laser-based beam shaping techniques [118].

While the proposed DEFT facility will occupy a length of around ten meters, an R&D effort currently underway at SLAC National Accelerator Laboratory seeks to reach an even more compact VHEE system using a cryogenic X-band accelerator to achieve VHEE beam energies in only one meter. This approach harnesses the enhanced efficiency and higher gradient obtainable in a distributed coupling linac, combined with the improved scaling at cryogenic temperatures [135] to reach the target gradient of 100 MeV/ m, already demonstrated in previous experiments at SLAC using a cryogenic X-band structure at the X-band Test Area (XTA), see Figure 16 [136]. A key aspect of the proposed VHEE system is the reliance on a commercial RF source, limiting the initial peak power to roughly 6 MW, but ensuring the final product can be widely adopted for commercial use. Advances in both the distributed coupling design and cryogenic operation of the normal conducting structure are necessary to maximize the power efficiency of the linac, allowing SLAC's one meter X-band structure to reach a gradient of 100 MeV/m while using a peak power of only 20 MW. A preliminary design of the cryogenic X-band linac is shown in Figure 17.

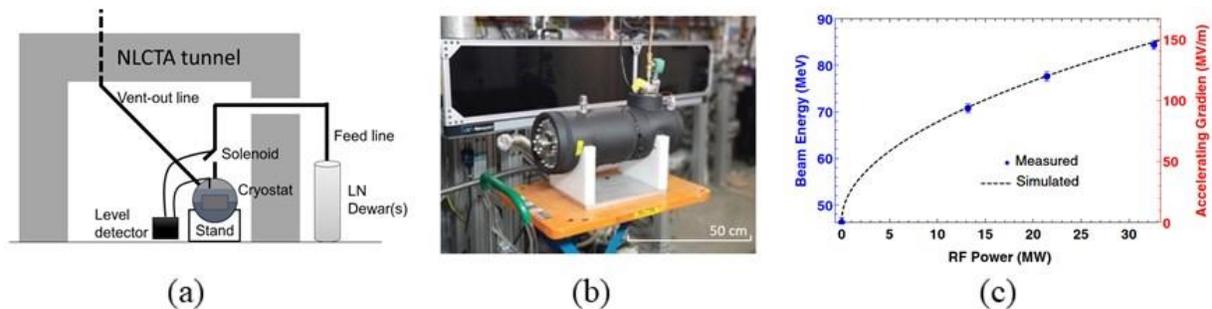

**Figure 16.** (a) Schematic of cryogenic test setup at XTA using liquid nitrogen. (b) Photo of linac and cryostat assembly prior to installation at XTA. (c) Measurements of accelerating gradient as a function of RF power from Ref. 137.

The SLAC system will rely on RF pulse compression to achieve a 20 MW peak power using the initial 6 MW pulse produced by the commercial klystron. RF pulse compression is a mature technology that has been used extensively for large scale accelerator applications to reconcile the need for short high-peak-power pulses with a cost-efficient long pulse, low power sources [137-139]. Recent advances in RF compressor technology have opened the door to compact structures that could dramatically reduce the system footprint while maintaining the capability to produce 4-fold pulse compression and isolate the source from the reflected RF signal from the cavities [140,141]. The SLAC VHEE program investigates multiple cavity designs, like the spherical cavity in Ref. 26, focusing on structures designed for high intrinsic quality factors, $Q_0$ up to 400,000, and high coupling factors, β up to 10. Active research in this area will continue to benefit efforts to design a new generation of compact, cost-efficient medical accelerators and the broader accelerator community that relies on pulse compressors to supply the peak powers needed for high gradient operation.

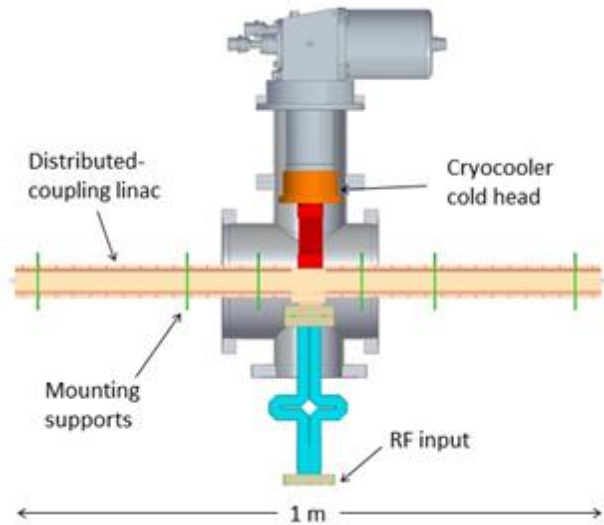

**Figure 17.** Schematic of the cryogenic X-band linac for VHEE under development at SLAC.

Proposed programs like the Cool Copper Collider (C$^3$) proposal [142] to realize an e$^+$e$^-$ collider for the study of the Higgs boson offer an exciting opportunity for synergistic research efforts on accelerator technologies, from the distributed coupling linac to the pulse compression system, which could enhance power efficiency for a single compact cancer therapy system up to a large-scale facility like C$^3$. In each case, the same underlying techniques are used to push the limits on achievable accelerating gradients with a cost-efficient system. Research from a C$^3$ linac R&D effort [143] would provide insight into features needed for a high gradient VHEE system and vice versa, including stringent performance reliability criteria optimized for substantial beam current under cryogenic operating conditions.

Fabrication of SLAC's X-band distributed coupling linac for VHEE will rely on a split-block approach which allows significant flexibility for CNC machining of the linac cavities and power coupling manifold into the copper slabs. This flexibility is critical for implementing a 135° phase advance linac design which further enhances the power efficiency, increasing the shunt impedance by nearly 10% compared to the $\pi$-mode. On-going collaboration with industry partners will facilitate the transition of SLAC's prototype VHEE system into modular industrialized equipment. Mass-production will be an important feature not only for commercialization generally but also for achieving FLASH capability with the VHEE system. In order to eliminate gantry motion and reach an ultra-high dose rate, the proposed FLASH VHEE system utilizes an array of 16 linacs, in the same architecture as the PHASER system for photon therapy [118]

VHEE beam energies as high as 250 MeV could be needed for treatment scenarios that use advanced techniques like spatially fractionated radiotherapy in combination with FLASH dose rates [144]. The demand for equipment that can deliver these beam energies on the meter scale has motivated a search for technology that can provide gradients exceeding 100 MeV/m at the FLASH dose rate. Accelerators operating at even higher frequencies than X-band, up into the mm-wave regime, offer an opportunity to provide VHEE FLASH-RT with unprecedented compactness. Efforts are already underway at SLAC National Accelerator Laboratory to apply recent advances in mm-wave linac fabrication and high-power testing (Figure 18) to the design of a VHEE accelerator operating at 94 GHz [145].

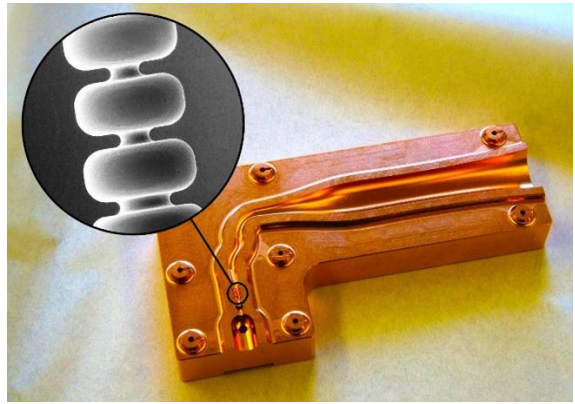

**Figure 18.** SLAC mm-wave linac prototype for high power testing.

The mm-wave and THz regime has long been frustratingly inaccessible for accelerator applications due to the absence of high-power sources and the challenges of implementing these small-scale structures. Motivated by the potential advantages of high frequency high gradient accelerators to explore the Energy Frontier, research has been conducted at SLAC on the fabrication of mm-wave accelerators [146] as well as methods of powering these structures and the physics of breakdown at these high frequencies [147]. New research is exploring the benefits of applying high gradient techniques like distributed coupling and cryogenic operation in the mm-wave regime [145,148]. These features offer the possibility for extremely power-efficient structures that could reach gradients of hundreds of MeV per meter. To power these structures, SLAC has partnered with the Air Force Research Laboratory to investigate active pulse compression at mm-wave frequencies, overcoming the limitations of low peak power available from commercial sources using techniques developed for nanosecond rf-power switching [149,150]. Active research and development is needed to push these structures beyond the few-cell prototypes that have undergone high power test to a full-scale system with demonstrated beam acceleration. A critical area of research, particularly in the pursuit of FLASH dose rates, will be the design of an electron gun compatible with these mm-wave structures [151].

### 3.6.3 Fast 3D high-speed beam scanner for hadron FLASH-RT

Proton therapy, and hadron therapy in general, allows potentially far greater dose shaping control than conventional photon therapy or VHEE through the energy-dependent Bragg peak, which determines the depth at which the peak dose will be delivered. Proton cyclotron facilities for cancer therapy routinely alter the proton beam energy used in treatment by passing the beam through a so-called range shifter, a physical barrier of material, typically plastic, which reduces the beam energy according to the thickness of the plate. While the strategy is a reliable and robust method for changing the beam energy, the process of switching between range shifter settings is time-consuming, on the order of a second [152] when compared to the desired time scale of a total FLASH treatment that is a few hundred milliseconds, and also degrades the lateral penumbra of the beam. Synchrotron facilities, used for both proton and carbon cancer treatment, can change the beam energy by adjusting the acceleration cycling settings, avoiding the mechanical motion of a range shifter, but face challenges to achieving the FLASH dose rate. The demand for high-speed changes to beam energy presents a tantalizing opportunity to apply accelerator technology in which RF-driven energy modulation could accomplish the same objective as the range shifter with changes on the sub-µs scale.

This research thrust has already gained traction in a program at SLAC National Accelerator Laboratory to develop a 3D high-speed beam scanner for hadron therapy. The objectives of this project are to design and demonstrate the component technology needed to modulate the beam energy and transverse steering, sufficient to cover a 4-liter volume at a FLASH dose rate. The energy modulator design builds on research concepts developed at SLAC for high energy physics applications, taking the high gradient capability of a distributed coupling S-band structure and using it to reach a +/- 30 MeV beam energy, equivalent to a range of 15 cm in treatment depth, in a one-meter structure [153].

SLAC's hadron scanning program tackles not only the challenge of RF-driven energy modulation but also transverse steering. Unlike conventional photon-based radiotherapy, VHEE and proton therapy allow for pencil beam scanning, which takes advantage of the Lorentz force to steer the charged particle trajectories. Thus far, transverse pencil beam scanning for protons has been routinely accomplished using electromagnets, which allow the beam to cover a large treatment field on the order of 30 cm by 40 cm at the patient isocenter [152,154]. This technique offers valuable flexibility in terms of coverage area with minimal beam distortion and, while not as fast as an RF-driven process, is compatible with the timescale of FLASH dose delivery. Changes to beam position can be accomplished on a few hundred microseconds scale [152]. Varian has already announced FLASH capability with their existing proton therapy equipment and has actively invested in FLASH therapy research through the FlashForward™ Consortium [155].

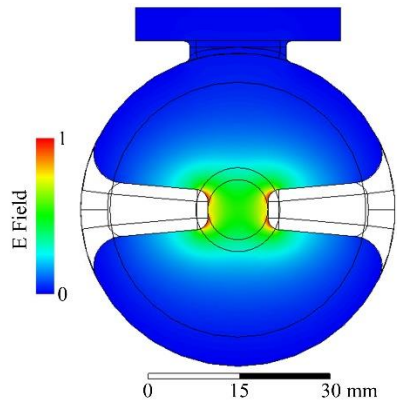

**Figure 19.** Electric field profile of the TE11-like mode shown in a cross-section of the deflector cell, with beam axis oriented into the page. The opposing posts (profiles shown in white) produce an RF dipole. Power is coupled in through the port at the top of the model simulated in ANSYS-HFSS.

These developments suggest that proton therapy will likely be one of the leading modalities for FLASH treatment in the near future. The ultrafast 3D beam shaping for hadron therapy championed by the SLAC-led collaboration on RF-driven beam manipulation offers an opportunity for the accelerator technology of the high energy physics community to revolutionize the speed and flexibility of proton therapy equipment. SLAC's proton deflector cavity, shown in Figure 19, is a prime example, taking inspiration from CERN's crab cavity research for beam steering [156] and optimizing a new cavity design for sub-relativistic protons [157]. The initial design for the SLAC 3D scanning system utilizes a few-cell deflector structure, with cavities oriented orthogonally for full range of transverse motion. The angular kick provided by the RF-driven deflector is enhanced by a set of static permanent magnet quadrupoles (PMQ). The effect of the PMQs will be to defocus in one plane and overfocus in the other. By compensating with the magnitude of the kick supplied by the deflector structure in each direction, this focusing action can be optimized for the maximum treatment area, covering around 15 cm x 22 cm for a proton beam energy of 200 MeV. This RF-driven approach to the proton beam modulation eliminates all mechanical motion and allows for ultra-fast switching between different energies and lateral positions.

The initial SLAC research has focused on the design and demonstration of prototypes of both the energy modulator and deflector, with high power testing underway at SLAC's facilities. In order to realize this technology in clinical settings, research and development will be needed to build the full-scale accelerator structures and conduct testing with a proton beam. SLAC has partnered with Electron Energy Corporation (EEC) to investigate designs for the PMQ system used to enhance the treatment field covered by the SLAC RF deflector. EEC's research on cryogenic PMQ designs offers unique advantages in terms of flexibility and performance reliability over a range of cryogenic temperatures, with applications not only to potential proton therapy equipment but also to programs pushing the Energy Frontier like the $C^3$ proposal to develop an $e^+e^-$ collider for the study of the Higgs boson [158].

**Acknowledgments** - This work was funded in part by the Department of Energy Contract DE-AC02-76SF00515.

## 4   Summary and Conclusions

FLASH radiation therapy (FLASH-RT) is the next frontier in radiation therapy for cancer. Initial preclinical and clinical research results look very promising, demonstrating enhanced protection of normal tissue, reduced toxicity and side effects, and good tumor control. Beam delivery questions remain, for example, the level of conformality or the use of multiple fields and intensity modulation, the role of the Bragg peak, possible limitations in tumor depth or size, and the importance of low-dose regions. For non-pulsed or CW accelerators, does the RF bunch create a FLASH effect equivalent to the pulse structure of clinical electron accelerators?

The research thrust into FLASH has already gained momentum at the National Accelerator Laboratories and leading universities in the United States. There are several programs to develop new compact solutions for FLASH-dose rate capable machines delivering X-rays, electrons, protons, and ions, with a comprehensive overview presented in this white paper. In addition to high-intensity accelerator R&D, precision beam control will need to be enhanced by a factor of 20-50 over a large field with corresponding and unprecedented ultra-fast detector response time and linearity. Investment into research and development and close collaboration between national accelerator labs, industry, and academic medical institutions will be required to realize and transfer the new technologies in clinical settings.